\documentclass[a4paper,fleqn,usenatbib]{mnras}

\usepackage[T1]{fontenc}
\usepackage{ae,aecompl}

\usepackage{graphicx}	% Including figure files
\usepackage{amsmath}	% Advanced maths commands
\usepackage{amssymb}	% Extra maths symbols

\title[Metallicities in M 83]{Metallicities of Young Massive Clusters in NGC 5236 (M83)
\thanks{Based on observations made with ESO telescopes at the La Silla Paranal Observatory under programme ID 085.B-0111(A)}}

\author[S. Hernandez et al.]{
Svea Hernandez,$^{1}$\thanks{E-mail: s.hernandez@astro.ru.nl}
S{\o}ren Larsen,$^{1}$
Scott Trager,$^{2}$
Lex Kaper,$^{3}$
and Paul Groot$^{1}$
\\
$^{1}$Department of Astrophysics / IMAPP, Radboud University, PO Box 9010, 6500 GL Nijmegen, The Netherlands\\
$^{2}$Kapteyn Astronomical Institute, University of Groningen, Postbus 800, NL-9700 AV Groningen, the Netherlands\\
$^{3}$Astronomical Institute Anton Pannekoek, Universiteit van Amsterdam, Postbus 94249, 1090 GE Amsterdam, The Netherlands
}

\date{Accepted 2017 September 12. Received 2017 September 11; in original form 2017 July 21}

\pubyear{2017}

\hypersetup{draft}

\begin{document}
\label{firstpage}
\pagerange{\pageref{firstpage}--\pageref{lastpage}}
\maketitle

\begin{abstract}
We present integrated-light spectra of 8 Young Massive Clusters (YMCs) in the metal-rich spiral galaxy NGC 5236 (M 83). The observations were taken with the X-Shooter spectrograph on the ESO Very Large Telescope. Through the use of theoretical isochrones and synthetic integrated-light (IL) spectra we derive metallicities and study the radial metallicity gradient observed through these young populations. For the inner regions of the galaxy we observe a relatively shallow metallicity gradient of  $-$0.37 $\pm$0.29 dex R$_{25}^{-1}$, agreeing with chemical evolution models with an absence of infall material and a relatively low mass loss due to winds in the inner parts of the disk.
We estimate a central metallicity of [$Z$] = $+$0.17 $\pm$ 0.12 dex, finding excellent agreement with that obtained via other methods (e.g. blue supergiants and J-band). We infer a metallicity of 12+log(O/H) = 8.75 $\pm$ 0.08 dex at R/R$_{25}$ = 0.4, which fits the stellar mass-metallicity relation (MZR) compilation of blue supergiants and IL studies. 
\end{abstract}


%%%%%%%%%%%%%%%%%%%%%%%%%%%%%%%%%%%%%%%%%%%%%%%%%%

%%%%%%%%%%%%%%%%% BODY OF PAPER %%%%%%%%%%%%%%%%%%

\section{Introduction}

The study of stellar chemical abundances has proven to be a strong tool in constraining the star formation histories of different galaxies, particularly our own Milky Way (MW; \citealt{wor98,mat03,ven04,pri05}). Knowledge of extragalactic chemical abundances is indispensable for understanding galaxy and chemical evolution on larger scales. \citet{leq79} and \citet{tre04}, amongst others, found that there is a correlation between the mass and metallicity of individual galaxies. This mass-metallicity relation (MZR) has been used to learn about star formation episodes, galactic winds and general chemical enrichment of star-forming galaxies \citep{mai08,kud12,fin08,lil13}. 
Furthermore, radial metallicity variations within a galaxy provide valuable information on the effects of merging, initial mass function,  infall and winds present in the galaxy \citep{pra00,san09,kud15,bre16}.  \par
Studies of the chemical evolution of galaxies have been limited by the difficulty in obtaining reliable abundances and metallicities. Extragalactic metallicities of star-forming galaxies are generally measured using H II region emission lines. Two types of analysis predominate in this field: ``strong line" and ``$T_{e}$-based". The former method is based on the ratio of fluxes from the the strongest forbidden lines relative to  H$\beta$ (typically O; \citealt{pag79}). On the other hand the ``$T_{e}$-based" method uses auroral lines to infer the electron temperature of the gas. Even though these lines are weaker across a wide range of metallicities, this method removes the dependence on ``strong line" calibrations \citep{rub94,lee04,sta05,and13}. One complication with this ``$T_{e}$-based" method occurs at metallicities close to solar and above, a regime where the auroral lines are extinguished \citep{sta05,bre05,erc10,zur12}. A well known problem with these two methods comes to light by comparing the metallicities inferred from the different diagnostics. Studies have observed that different methods yield obvious systematic offsets in the inferred metallicities \citep{ken03,kew08,mou10,lop12}. However, even with its metallicity range limitations, \citet{sta05} predicts that the ``$T_{e}$-based" method provides more robust measurements below solar metallicities. \par

In the last decade spectroscopic observations of both red (RSG) and blue (BSG) supergiants have become an important tool to study the metallicities of extragalactic populations. The supergiant technique has been used as an alternative method for measuring metallicities and abundance gradients beyond the Milky Way and even the Local Group \citep{bre06,eva07,kud13, dav10, gaz14, lar15, kud16, bre16}. Results from this technique show excellent agreement with abundances obtained from the ``$T_{e}$-based" method \citep{kud12,kud13, kud14,hos14,gaz15}.\par

In addition to spectroscopic observations of H II regions and supergiants, other studies have developed techniques to obtain detailed abundances from high resolution (\textit{R} $\sim$25,000) spectroscopic observations of unresolved extragalactic globular clusters (GC; \citealt{mc08,mc02,ber05,col09,col11,col12,lar12,lar14}). With similar masses as GCs (> 10$^{4}$$M_{\odot}$), young massive clusters (YMCs) are characterised by their young ages (< 100 Myr; \citealt{por10}). The identification of significant populations of YMCs in galaxies with on-going star formation \citep{lar992,lar04} has allowed the study of star formation histories and chemical evolution of individual galaxies to expand its parameter space. It is now feasible to learn about the recent chemical evolution of the stellar components in extragalactic environments. In \citet{her17} we demonstrate that detailed abundance analysis is possible for intermediate-resolution observations (\textit{R}< 8,800) of YMCs using NGC1313  ($\sim$4 Mpc) and NGC1705 ($\sim$5 Mpc) as test cases applying the spectral synthesis technique. Furthermore, the \textit{J}-band method was recently used to measure accurate metallicities of extragalactic YMCs \citep{gaz14, gaz14_2, lar15}. An additional advantage of studying the chemical histories of galaxies using star clusters (GCs/YMCs) over H II regions is the fact that H II regions trace the present-day metallicity of the gas phase, while star clusters can provide information on a broad range of ages/times. This paper aims to further exploit the recently developed techniques for integrated-light studies by exploring higher metallicity environments (above solar), such as those observed in the spiral galaxy NGC 5236 (M 83; \citealt{bre02, bre05}) located at a distance of 4.9 Mpc \citep{jac09}. \par

In this work we present the analysis of intermediate-resolution integrated-light observations of 8 YMCs distributed throughout NGC 5236 in an effort to determine the metallicity gradient across the disk of the galaxy. In Section ~\ref{sec:obs} we provide a brief description of the X-Shooter spectrograph, target selection and science observations, followed by details on our data reduction approach. In Section ~\ref{ana} we present the abundance analysis applied in this work where we include information on the atmospheric models, stellar parameters, and creation of the synthetic observations. We introduce our main results in Section ~\ref{results} followed by our discussion in Section ~\ref{discuss}. We summarise our main remarks in Section ~\ref{con}. 
%__________________________________________________________________

\section{Observations and Data Reduction}
\label{sec:obs}
\subsection{Instrument, Target Selection and Science Observations} 
\label{InsTagObs}
The data analysed here were taken with the X-Shooter spectrograph on ESO's Very Large Telescope (VLT), located on Cerro Paranal, Chile \citep{ver11}. The instrument has a wavelength coverage between 3000-24800\r{A}. This broad coverage is possible due to its three-arm system, UV-Blue (UVB), Visible (VIS), and Near-IR (NIR). Depending on the configuration the spectrograph observes at resolutions ranging from R= 3000 to 17000. The science exposures use slit widths 1.0", 0.9" and 0.9" providing resolutions of R $\sim$ 5100, 8800 and 5100 for the UVB, VIS and NIR arms, respectively. The data was collected using the standard nodding mode with an ABBA sequence under GTO program 085.B-0111A in April 2010. Telluric standard stars were observed as part of the GTO program. Flux standard star observations were collected through the ESO X-Shooter calibration program and dowloaded from the archive to be used in the reduction of the science exposures. Due to the low signal-to-noise (S/N) in the NIR exposures, this work makes use of the science observations obtained with the UVB and VIS arms only. In Table~\ref{table:obs} we list the different cluster IDs, coordinates, exposure times, S/N values for the corresponding arms and the seeing. \par
\begin{figure}
   \resizebox{\hsize}{!}
            {\includegraphics[width=11.2cm]{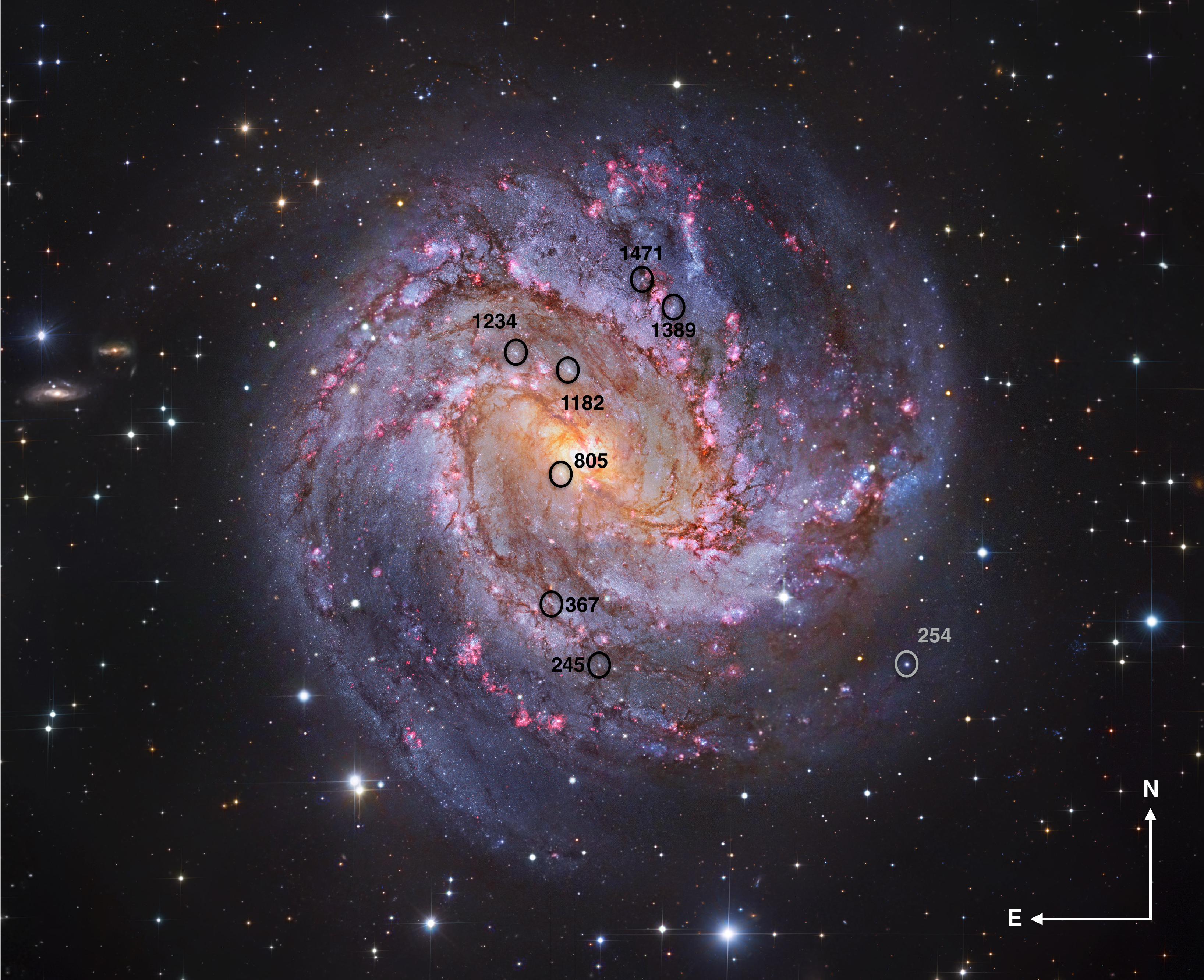}}
      \caption{A colour-composite image of NGC 5236 observed with the 8.2-meter Subaru Telescope (NAOJ), the 2.2-meter Max Planck-ESO telescope and the Hubble Space Telescope. We mark the location of the different YMCs studied as part of this work. Image Credit: Subaru Telescope (NAOJ), Hubble Space Telescope, European Southern Observatory. Processing and Copyright: Robert Gendler. }
         \label{Fig:galaxy}
   \end{figure}

The YMCs were selected using the catalog by \citet{lar04}. The selection criteria required uncontaminated objects and magnitudes brighter than V=19. In Figure~\ref{Fig:galaxy} we show the location of the individual YMCs in NGC 5236 analysed in this work. 

\begin{table*}
\caption{X-Shooter Observations}
\label{table:obs}
\centering 
\begin{tabular}{cccccccc} 
 \hline \hline\\
Cluster & RA & DEC &  \multicolumn{2}{c}{$t_{exp}$ (s)}  &  \multicolumn{2}{c}{S/N (pix$^{-1}$)}& Seeing ($\rm \arcsec$) \\
 & (J2000) & (J2000) & UVB & VIS & UVB & VIS & \\
 \hline \\
NGC 5236-245 & 204.248735 & $-$29.91319	 & 2000.0 & 1980.0 & 12.0 & 10.6 & 1.0\\
NGC 5236-254 & 204.167693 & $-$29.91352	 & 2000.0 & 1980.0  & 17.0 & 15.4 & 0.8\\
NGC 5236-367 & 204.261056 & $-$29.89977	 & 2000.0 & 1980.0  & 16.5 & 13.8  & 0.7\\
NGC 5236-805 & 204.258142 & $-$29.86966	 & 1620.0 & 1600.0  & 53.4 & 33.4  & 0.6\\
NGC 5236-1182 & 204.255822 & $-$29.84655 & 2000.0 & 1980.0  & 58.5 & 35.4  & 0.7\\
NGC 5236-1234 & 204.270055 & $-$29.84329 & 2000.0 & 1980.0  & 26.8 & 14.8 & 0.6\\
NGC 5236-1389	 & 204.228690 & $-$29.83262 & 2000.0 & 1980.0 & 27.8 & 17.1 & 0.6\\
NGC 5236-1471	 & 204.236777 & $-$29.82624 & 2000.0 & 1980.0 & 18.2 & 14.9 & 0.9\\
 \hline
\end{tabular}
\end{table*}

\subsection{Data Reduction}\label{DR}
The basic reduction steps are performed using the standard ESO Recipe Execution Tool (EsoRex) v3.11.1 and the public release of the X-Shooter pipeline v2.5.2. The spectral extraction is done using the IDL algorithms developed by \citet{che14}. \par
We flux calibrate the data using exposures of Feige 110, a spectrophotometric standard object observed close in time to the science data. For a more detailed discussion on the individual steps involved in the flux and telluric corrections we point the reader to \citet{her17}. Briefly summarised, we create response curves for each of the science frames where we correct for exposure time and atmospheric extinction. For these response curves we use the same flat field and master bias frames applied to the corresponding science exposures. The telluric corrections for the VIS exposures are done using the telluric library compiled by the X-Shooter Spectral Library (XSSL) team along with a Principal Component Analysis (PCA) routine created by \citet{che14}. This PCA algorithm removes and reconstructs the strongest telluric absorptions. 

%__________________________________________________________________

\section{Abundance Analysis}\label{ana}
We make use of the analysis method developed by \citealt{lar12} (hereafter L12) to obtain detailed abundances from integrated-light observations of star clusters. The L12 method was originally designed and tested using high-dispersion (\textit{R} $\sim$ 40,000) spectroscopic observations, and extended to intermediate-resolution (\textit{R} < 8,800) observations by \citet{her17}. \par
Briefly summarised, we create a series of high-resolution (\textit{R} $\sim$ 500,000) simple stellar population (SSP) models where we include every evolutionary stage present in the star cluster. First a series of atmospheric models is created using ATLAS9 \citep{kur70} and MARCS \citep{gusta08}. The former are used for stars with T$_\mathrm{eff}$ > 5000 K, and the latter for T$_\mathrm{eff}$ < 5000 K. Synthetic spectra for individual stars are created using SYNTHE \citep{kur79,kur81} and TURBOSPECTRUM \citep{plez12} for ATLAS9 and MARCS models respectively. The spectra are then co-added to generate a synthetic integrated-light spectrum for the star cluster in question. The synthetic spectra are then compared to the X-Shooter observations and the abundances are modified until the best match (minimum $\chi^{2}$) between model and observations is obtained. \par

In this work we make use of a scaling parameter relative to Solar composition and apply it to all of the specified abundances. We note that the metallicity [Z] derived from this analysis is a measure of the integrated abundances of different chemical elements, including, and not limited to, $\alpha$- and Fe-peak elements. The current software uses Solar composition from \citet{gre98}. Additionally, the code allows the user to assign weights to different parts of the spectrum on a pixel-to-pixel basis, with values ranging from 0.0 (exclusion) to 1.0 (inclusion). For our analysis we set the weights to 0.0 in regions affected by instrumental features, telluric contamination and nebular/ISM emission. 

\subsection{Stellar Parameters} \label{stel_par}
We create a Hertzsprung-Russell diagram (HRD) to cover and represent every evolutionary stage in the YMC using the theoretical models from PARSEC v.1.2S \citep{bre12}. Previous studies have found the metallicity of the disc of NGC 5236 to be above Solar \citep{bre02}. For the initial selection of the isochrones we adopt a metallicity [Z] = 0.3 and YMC ages found in the literature. The cluster ages have been estimated from photometric observations and applying the S-sequence age calibration defined by \citet{gir95}. This method relies on an age sequence derived from fitting the average colours of bright LMC star clusters in the U-B vs B-V space and has been applied to star clusters external to the LMC \citep{bre96}. Typical errors on these photometric ages are a factor of 2.
 In Table ~\ref{table:prop} we show the YMC properties, including the ages, masses, normalised galactocentric distance, and effective radii, along with their corresponding literature reference. \par
The stellar parameters (T$_\mathrm{eff}$, log $g$, M) are extracted from these theoretical isochrones assuming an Initial Mass Function (IMF) following a power law, $dN/dM \propto M^{-\alpha}$, adopting a \citet{sal55} exponent of $\alpha$ = 2.35, and a lower mass limit of 0.4 $M_{\odot}$.\par

An additional feature in the L12 code is the capability to fit for the microturbulent velocity, $v_{t}$. We initially fit for [Z] and $v_{t}$ simultaneously for all 8 YMCs. The code fits for a single $v_{t}$ value and applies it to all the stars in the cluster, irrespective of type. We find a poorly constrained mean microturbulence of $\langle \nu_\mathrm{t} \rangle = 2$ $\pm$ 1km s$^{-1}$. Due to the large uncertainties in the calculated $v_{t}$ we perform several tests changing the $v_{t}$ from 1 to 2 km s$^{-1}$ for stars with T$_\mathrm{eff}$ < 6000 K. Changing the $v_{t}$ values from 1 km s$^{-1}$ to 2 km s$^{-1}$ changes the overall metallicity on average by $\lesssim$0.1 dex, with the exception of NGC 5236-1471 where [Z] changes by 0.19 dex. For the rest of our analysis we adopt the following microturbulent values: $v_{t}$  = 2 km s$^{-1}$ for stars with T$_\mathrm{eff}$ < 6000 K, $v_{t}$ = 4 km s$^{-1}$ for stars with 6000 < T$_\mathrm{eff}$ < 22000 K \citep{lyu04} and $v_{t}$ = 8 km s$^{-1}$ for stars with  T$_\mathrm{eff}$ > 22000 K \citep{lyu04}, similar to what was used in \citet{her17}.

\begin{table}
\caption{Young Massive Cluster Properties. References for each of the clusters are listed as footnotes.}
\label{table:prop}
 \centering 
\begin{tabular}{ccccc} 
 \hline  \hline \
Cluster & log(age) & Mass$_{\rm phot}$ & $R/R_\mathrm{25}$&R$_{\rm eff}$ \\
 & & [$M_{\odot}$] & & (pc)\\
  \hline\\ 
 NGC 5236-245$^a$&$ 8.00$& $1.4 \times 10^5$&0.44 &5.4 \\
 NGC 5236-254$^b$&$ 8.25$& $2.7 \times 10^5$&0.91 & 10.1\\
 NGC 5236-367$^a$&$ 7.85$& $1.1 \times 10^5$&0.32 & 4.7\\
 NGC 5236-805$^c$&$ 7.10$& $2.0 \times 10^5$&0.05 & 2.3\\
NGC 5236-1182$^d$&$ 7.45$& $2.1 \times 10^5$&0.17 & 6.8\\
NGC 5236-1234$^d$&$7.45$&$8.1 \times 10^4$&0.26 & 7.2\\
NGC 5236-1389$^e$&$ 7.69$& $1.1 \times 10^4$&0.39 & 8.7\\
NGC 5236-1471$^a$&$ 7.76$& $8.7 \times 10^4$&0.40 & 2.9\\
 \hline 
 \end{tabular}
\\ \textsuperscript{$a$} \citet{lar04,lar09,bas13},\\ \textsuperscript{$b$} \citet{lar06},
\textsuperscript{$c$} \citet{larr04},\\\textsuperscript{$d$}\citet{lar11}, 
\textsuperscript{$e$} \citet{lar99} \\
\end{table}

\subsection{Instrumental Resolution and Velocity Dispersion} \label{resol}
As mentioned before, we create a high resolution ($R \sim$ 500,000) model spectra which we degrade to match the resolution of our science observations. The L12 code has the option of fitting for the best Gaussian dispersion value ($\sigma_{sm}$) used to smooth the model spectra. Using this feature we fit for the best $\sigma_{sm}$ and [Z] values, analysing 200 \r{A} of data at a time. We repeat this procedure to obtain the $\sigma_{sm}$ of each of the YMCs in our sample. In general the $\sigma_{sm}$ accounts for the finite instrumental resolution ($\sigma_\mathrm{inst}$) and the internal velocity dispersions in the cluster ($\sigma_\mathrm{1D}$).  \par

\citet{che14} report that the X-Shooter resolution in the UVB arm varies with wavelength, but remains constant in the VIS arm. Following this same assumption and that where the resolving power represents a Gaussian FWHM, we use the same instrumental resolution as that presented in \citet{her17}, $\sigma_\mathrm{inst}$ = 14.47 km s$^{-1}$. We estimate the cluster velocity dispersions using the average $\sigma_{sm}$ calculated from the VIS observations alone. The line-of-sight velocity dispersion for each of the clusters is obtained through the following relation,

\begin{equation}
\sigma_\mathrm{1D} = \sqrt{ \sigma_{sm}^{2} - \sigma_\mathrm{inst}^{2}}
\end{equation}

Our analysis assumes an instrumental resolution set by the slit width alone. We note that the velocity dispersions may be underestimated if the actual resolution is higher than the standard instrumental resolution (e.g. if the seeing FWHM is smaller than the slit width). In Table ~\ref{table:derived} we summarise the derived line-of-sight velocity dispersions for the different YMCs included in this work. We note that for YMC NGC 5236-805 \citet{larr04} infer a line-of-sight velocity dispersion of $\sigma_\mathrm{1D}$ = 8.1 $\pm$ 0.2 km s$^{-1}$ which is comparable to our measured velocity dispersion of $\sigma_\mathrm{1D}$ = 7.7 $\pm$ 4.4 km s$^{-1}$. \par
\begin{figure}
    \resizebox{\hsize}{!}
            {\includegraphics[width=11.2cm]{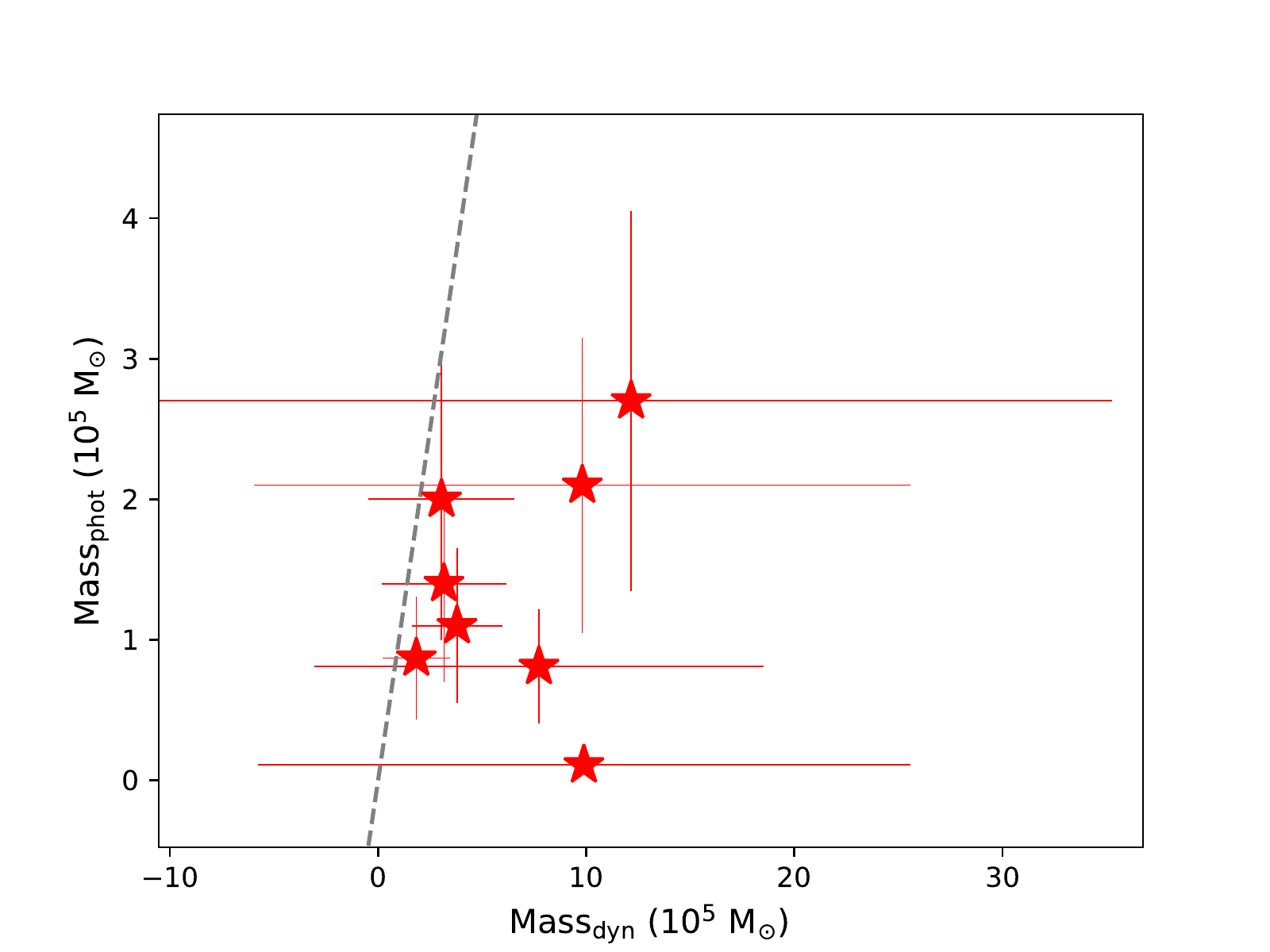}}
      \caption{Photometric mass, Mass$_{\rm phot}$ as a function of dynamical mass, Mass$_{\rm don}$. We note that the errors on the photometric masses are a factor of 2 \citep{bas13}. In grey we show the line of equal value. }
         \label{mass_mass}
   \end{figure}

We take the  $\sigma_\mathrm{1D}$ along with the effective radii listed in Table ~\ref{table:prop} and estimate the dynamical masses (Mass$_{\rm dyn}$) using the following relation

\begin{equation}
M_{\rm dyn} = \alpha \; \frac{\sigma_{\rm 1D}^{2} \;R_{\rm eff}}{G}
\end{equation}

\noindent where $\alpha \sim 9.75$. The cluster masses listed in Table ~\ref{table:prop}, Mass$_{\rm phot}$, are estimated using the $M/L$ models of Bruzual \& Charlot using a Salpeter IMF. In Figure ~\ref{mass_mass} we show the  Mass$_{\rm phot}$ as a function of Mass$_{\rm dyn}$. The dynamical masses appear to be slightly higher than the photometric masses, however both are consistent within the large uncertainties.

%
%________________________________________________________________
\section{Results}\label{results}
\begin{figure*}
   %\resizebox{\hsize}{}
    \centering
            {\includegraphics[scale=0.80]{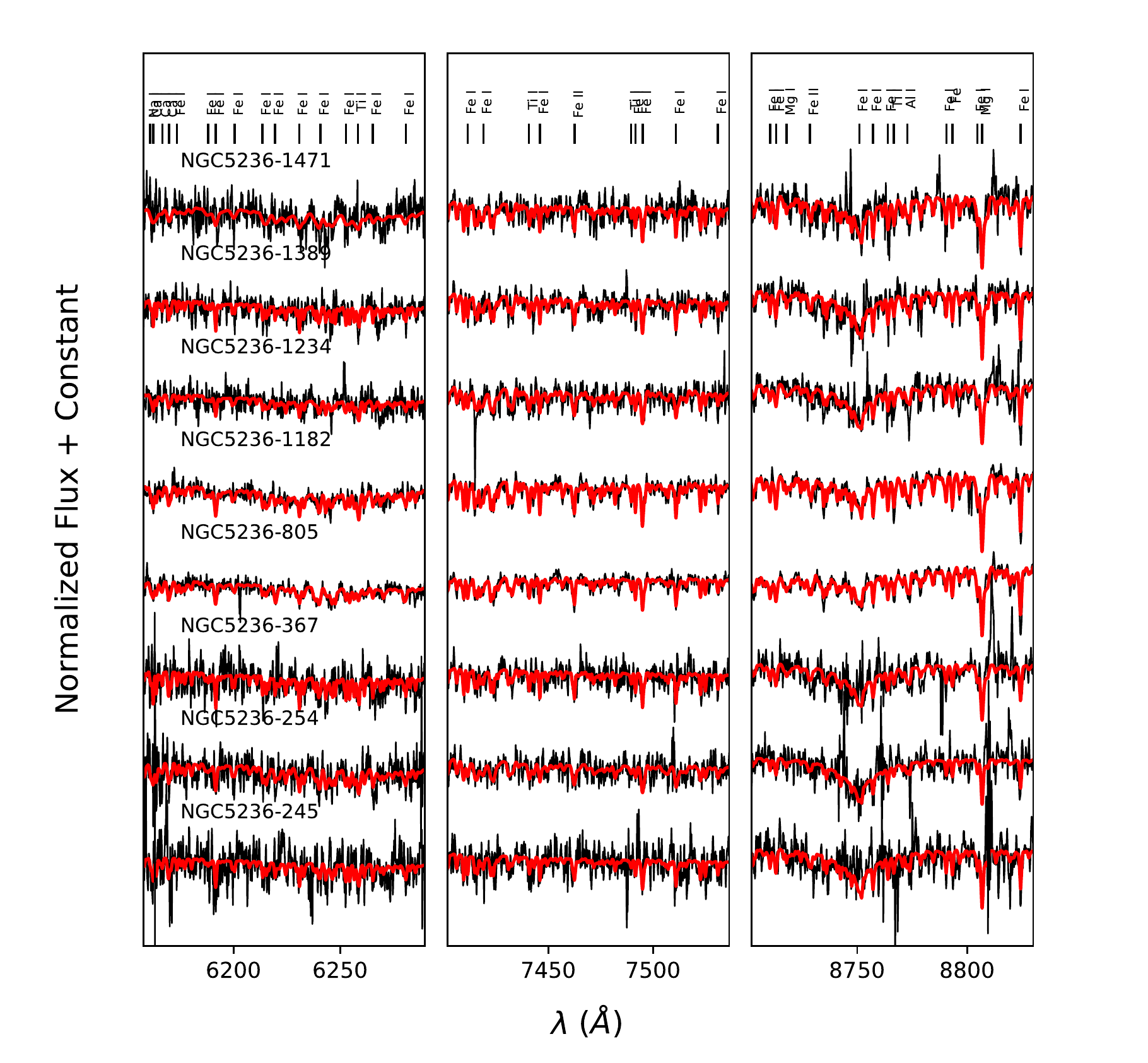}}
      \caption{Normalised integrated spectra for individual observations (in black) with its corresponding best fitting models (in red). We have added a constant offset for the benefit of visualisation. The cluster IDs are shown.}
         \label{Fig:SpecALLMod}
   \end{figure*}

After obtaining the best smoothing parameter ($\sigma_{sm}$), we proceed to estimate [Z], keeping $\sigma_{sm}$ fixed. Similar to the analysis in \citet{her17} we fit for the [Z] of each of the clusters scanning the UVB and VIS wavelengths using 200 \r{A} bins, excluding telluric-contaminated bins and those affected by the noise near the edge of the arms (5200-5400 \r{A} and 5400-5600 \r{A} for the UVB and VIS arms respectively). We use a cubic spline with 3 knots to match the model continua to the observed spectra. In Figure ~\ref{Fig:SpecALLMod} we show example synthesis fits for all the YMCs. The individual metallicity measurements obtained for the different wavelength bins and their corresponding 1-$\sigma$ uncertainties from the $\chi^{2}$ fit are listed in Tables ~\ref{z245} through ~\ref{table:z1471} of the Appendix. Once the minimum $\chi^2$ ($\chi_{\rm min}^{2}$) has been found the 1-$\sigma$ uncertainties are estimated by varying the metallicity until $\chi^{2} = \chi_{\rm min}^{2} + 1$. We note that in our final bin consideration we exclude UVB wavelengths  between 4400-5200 \r{A} mainly because in every iteration when we change the input isochrone (different age and metallicity), the measured [Z] values for these wavelengths change drastically, this in contrast to the rest of the bins where the values remain relatively constant in spite of a change in input isochrone. These changes were in the order of $\sim$ 0.2-0.4 dex, depending on the cluster. This behaviour was observed in all YMCs. Given the broad wavelength coverage in X-Shooter data, the exclusion of these bins does not impact our analysis.\par
In Table ~\ref{table:derived} we present weighted averaged metallicities, their corresponding errors ($\sigma_{\rm err}$), the number of bins ($N$) included in the analysis and the estimated radial velocities ($v_{\rm rv}$). The $\sigma_{\rm err}$ is calculated using Eq. (5) of ~\citet{her17}, where we account for the number of individual measurements ($N$) when estimating the errors on the mean metallicities along with the weighted standard deviation, 
\begin{equation} \label{eq:STD}
\sigma_\mathrm{STD} =  \sqrt{\frac{ \sum_{}^{} w_{i} \; ( Z_{i} - \bar{Z_{w}})^{2}}{\frac{N_{\rm nonz} - 1}{N_{\rm nonz}} \; \sum_{}^{} w_{i}}}.
\end{equation}
In Eq. \ref{eq:STD} the individual weights are represented by $w_{i}$ and defined as $w_{i} = 1/ \sigma_{i}^{2}$, $N_{nonz}$ is the number of non-zero weights, the different bin metallicities are identified as $Z_{i}$ and the weighted average metallicities as $\bar{Z}_{w}$. This approach for $\sigma_{\rm err}$ is chosen given that the scatter in individual measurements is larger than the errors based on the $\chi^{2}$ fitting, therefore more representative of the actual uncertainties in the measurements. \par
\begin{figure}
   \resizebox{\hsize}{!}
            {\includegraphics[width=11.2cm]{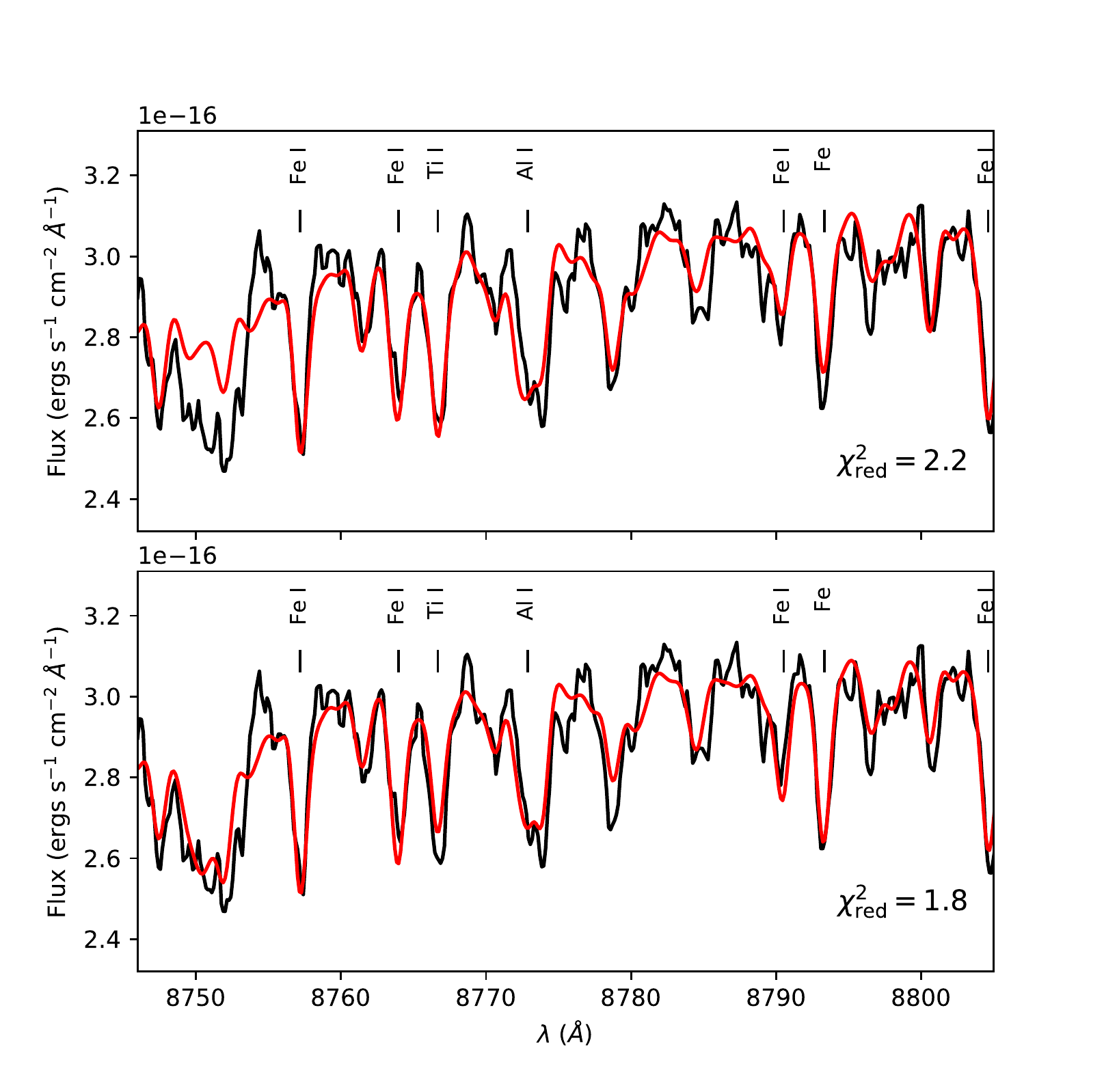}}
      \caption{In black we show the X-Shooter observation of NGC 5236-805. Top: In red we display the best model spectrum for NGC 5236-805 generated with an isochrone of log(age)=7.1 and [Z] = $+$0.33 dex. Bottom: In red is the model spectrum for the same YMC using isochrone of log(age)=7.1 and [Z] = $+$0.20 dex. We show the final $\chi^{2}_{\rm red}$ in the corresponding panels. See text for details. }
         \label{fig:ymc805}
   \end{figure}

In \citet{her17} we observed that selecting an isochrone to self-consistently match the inferred metallicity for the youngest YMC with a log(age)= 7.1, NGC1705-1, does not necessarily converge on the best model spectrum in spite of measuring similar metallicities (See Figure 7 in \citealt{her17}). We note that the behaviour seen in NGC1705-1 was not present in the analysis of the youngest cluster in this study or in any of the other YMCs. Using an initial isochrone of metallicity [Z] $\sim+$ 0.33 dex for NGC 5236-805, we estimate an overall metallicity of [Z] $\sim+$ 0.13 dex. We then continue our analysis changing the input isochrone metallicity to [Z] $\sim+$ 0.20 dex, and derive a final metallicity of [Z] $\sim+$ 0.17 dex. In contrast to NGC1705-1, visually inspecting the individual fits shows that the best model spectra generated using the isochrones with metallicity similar to the derived values match the observations well and decreases the final $\chi^{2}_{\rm red}$ values (see Figure ~\ref{fig:ymc805}).

\begin{table}
\caption{Young Massive Clusters derived quantities.}
\label{table:derived}
 \centering 
\begin{tabular}{cccccc} 
 \hline  \hline \
Cluster & $\sigma_\mathrm{1D}$ & [Z] & $\sigma_\mathrm{err}$& $N$& $v_{\rm rv}$\\
 & km s$^{-1}$ &  dex & dex & & km s$^{-1}$\\
  \hline\\ 
 NGC 5236-245& 5.1$\pm$2.4& $+$0.02& 0.06&9 & 559$\pm$5\\
 NGC 5236-254&7.3$\pm$6.9& $-$0.14& 0.11&9 & 558$\pm$32\\
 NGC 5236-367&6.0$\pm$1.7& $+$0.00& 0.09&9 & 535$\pm$5\\
 NGC 5236-805&7.7$\pm$4.4& $+$0.17& 0.12&9 & 496$\pm$4\\
NGC 5236-1182&8.0$\pm$6.4& $+$0.17& 0.13&9 & 461$\pm$5\\
NGC 5236-1234&6.9$\pm$4.8& $+$0.06& 0.21&9 & 441$\pm$4\\
NGC 5236-1389&7.1$\pm$5.6& $+$0.04 & 0.09&9 & 472$\pm$3\\
NGC 5236-1471&5.3$\pm$2.3& $+$0.12& 0.09&9 & 469$\pm$2\\
 \hline 
 \end{tabular}
\end{table}

\subsection{Sensitivity to ATLAS9/MARCS models and spectral synthesis computations } \label{sensitivity}
\begin{figure*}
   \resizebox{\hsize}{!}
            {\includegraphics[width=11.2cm]{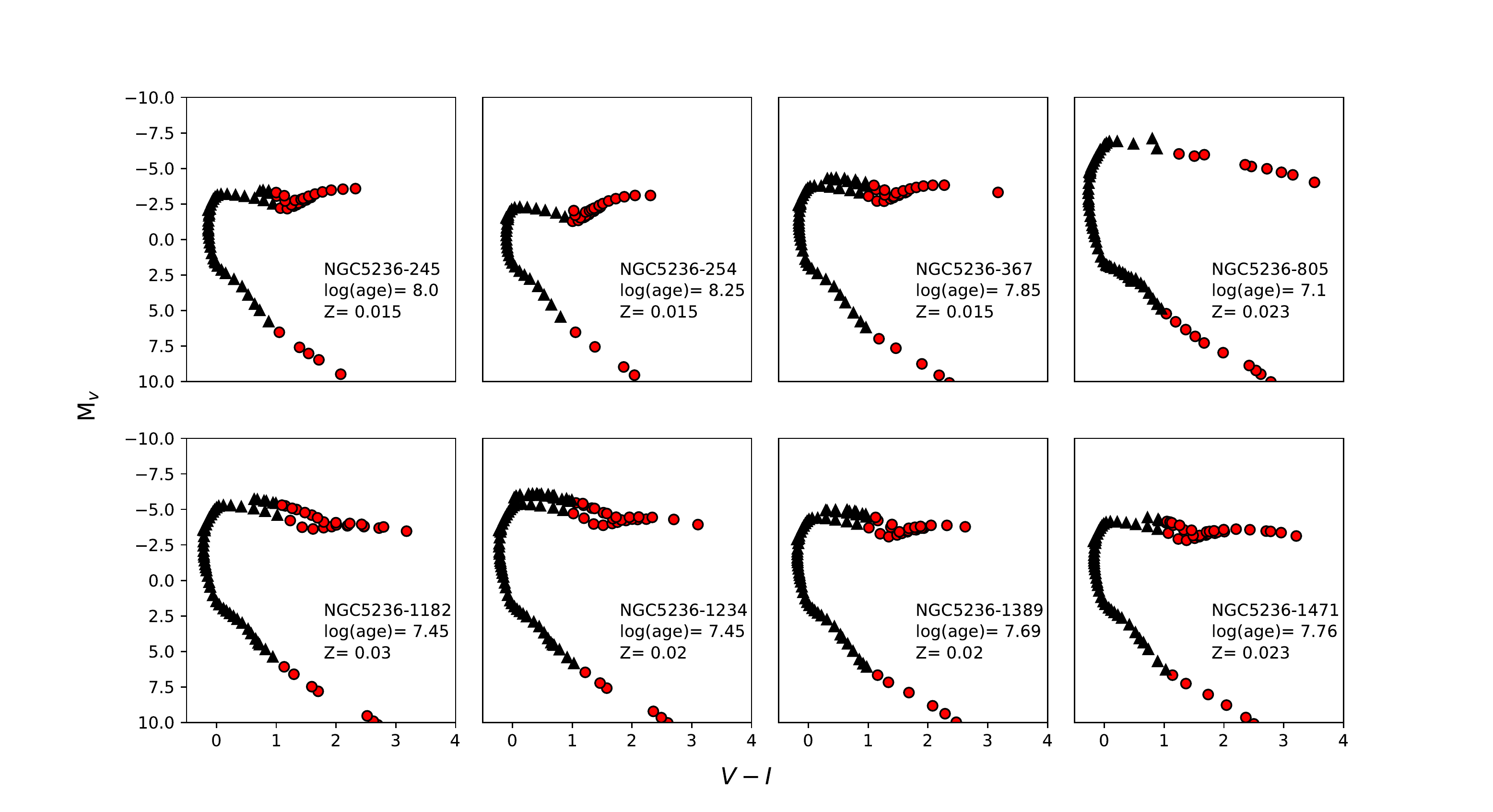}}
      \caption{Theoretical isochrones corresponding to the best fitting metallicities, Z, where Z$_{\odot}$ = 0.017 \citep{gre98}. In red circles we show stars with T$_\mathrm{eff}$ < 5000 K, for which we use MARCS models. In black triangles we show those stars with warmer temperatures (T$_\mathrm{eff}$ > 5000 K) for which we used ATLAS9 models.}
         \label{Fig:teff_iso}
   \end{figure*}

As described in Section ~\ref{ana}, we use two different sets of models depending on the T$_\mathrm{eff}$ of the star. For cool stars (T$_\mathrm{eff}$ < 5000 K), we use MARCS atmospheric models along with the TURBOSPECTRUM software to compute the synthetic spectra. The MARCS models allow for spherically-symmetric stellar atmospheres, generally preferred for stars with extended atmospheres compared to the plane-parallel symmetry used in ATLAS9 models. In this work we use a boundary in T$_\mathrm{eff}$ to separate the majority of giants from dwarfs. In Figure ~\ref{Fig:teff_iso} we show the final isochrones used for the different clusters. Displayed in red circles are those stars with T$_\mathrm{eff}$ < 5000 K, mainly covering the giant-like types. We point out that some lower main sequence stars are also identified to have T$_\mathrm{eff}$ < 5000 K, however, their contribution to the integrated-light spectrum is rather small. Stars with T$_\mathrm{eff}$ > 5000 K are shown in black triangles. From Figure ~\ref{Fig:teff_iso} is clear that an T$_\mathrm{eff}$ boundary of 5000 K reasonably covers the supergiant regime, located in the evolved branch of the HRD (red circles with M$_{\rm V} \sim -2.5$). \par

To explore how sensitive our metallicity measurements are to the different model choices, we compare the metallicities inferred using different T$_\mathrm{eff}$ boundaries. In the first run we set an T$_\mathrm{eff}$ boundary of 3500 K. With this temperature boundary we use ATLAS9 models for the majority of the stars, including giants. The second run uses a boundary of T$_\mathrm{eff}$  = 5000 K. Given the ages and metallicities of the different clusters, the second run uses MARCS models for most of the giants (See Figure ~\ref{Fig:teff_iso}). The results of this study are presented in the second column of Table ~\ref{table:soft}. Changing the T$_\mathrm{eff}$ boundary from 3500 K to 5000 K varies the inferred metallicity as much as 0.26 dex in the most extreme case. We note that in most cases this change in T$_\mathrm{eff}$ modifies the measured [Z] by $<$ 0.10 dex. \par

Given the intrinsic dependence of our analysis on the selection of theoretical models we investigate how sensitive our results are to the input isochrone ages. We recalculate the metallicities of each of the clusters modifying the ages by a factor of 2$\times$. In the third column of Table ~\ref{table:soft} we show the results of this comparison. Changing the input ages by 2$\times\log$(age) we see that the average metallicity change amongst all 8 YMCs is $\sim$0.1 dex, with the highest metallicity change seen for NGC5236-805 with a difference of $\Delta$ [$Z$] = $-$0.20. 

We point out that the work presented here is based solely on LTE models. At this moment we do not correct for any non-LTE (NLTE) effects. Such corrections are dependent on the physical parameters of each individual star, which makes NLTE corrections particularly complicated for integrated-light analysis. In the case of RSGs, studies have estimated NLTE corrections for [Fe/H] abundances in the order of $\sim$0.1 dex or lower \citep{ber12}. Higher NLTE corrections have also been predicted for some $\alpha$-elements with values ranging from $-0.4$ to $-0.1$ dex \citep{ber15}. 

\begin{table}
\caption{Sensitivity to ATLAS9/MARCS models}
\label{table:soft}
 \centering 
\begin{tabular}{ccc} 
 \hline  \hline \
Cluster & $\Delta$ T$_{\rm eff}$& $\Delta$ t\\
& $+$1500 K & $2\times \log$(age) \\
  \hline\\ 
 NGC 5236-245& $-$0.01& $-$0.12\\
 NGC 5236-254& $+$0.09&$-$0.01\\
 NGC 5236-367& $-$0.23& $+$0.02\\
 NGC 5236-805& $-$0.09& $+$0.20\\
NGC 5236-1182& $-$0.01& $+$0.04\\
NGC 5236-1234& $-$0.26& $-$0.05\\
NGC 5236-1389& $+$0.15& $-$0.14\\
NGC 5236-1471& $+$0.08& $+$0.06\\
 \hline 
 \end{tabular}
\end{table}

%\subsection{Emission Lines} \label{emission}
%\textbf{Similar to what was observed in both YMCs in \citet{her17} all of the systems analysed in this work display strong Balmer emission lines, mainly in H$\alpha$. The H$\alpha$ profiles vary from cluster to cluster, with some exhibiting P Cygni profiles. One possible cause for these emission lines is the presence of Be stars. These type of stars are identified mainly by their rotational velocities usually in the order of several hundred km s$^{-1}$ and are commonly found in young clusters with ages $<$100 Myr \citep{mcs05, wis06, mat08}.  We note that the these same clusters also exhibit strong emission in other nebular lines, such as} $\ion{N}{ii}$, \ion{O}{ii}, and \ion{S}{ii}, which could hint at the presence of gas.  \par

%   \begin{figure*}
  % \resizebox{\hsize}{!}
    %        {\includegraphics[width=11.2cm]{Halpha.pdf}}
     % \caption{X-Shooter observations of the H$\alpha$ profiles for the different YMCs studied in this work. The different panels display the corresponding cluster ID.  }
%         \label{Fig:balmer}
  % \end{figure*}

%________________________________________________________________
\section{Discussion}\label{discuss}
\subsection{Mass-Metallicity Relation} \label{mzr}
The mass-metallicity relation (MZR) is an important diagnostic tool in the inference of star formation scenarios, galactic winds and chemical histories of galaxies. As mentioned earlier this relationship was observed in star-forming galaxies by ~\citet{leq79} through the study of H II regions in irregular and blue compact galaxies.  ~\citet{tre04} and ~\citet{and13} later expanded this study by analysing $\sim$53,000 and $\sim$200,000 star-forming galaxies and their gas-phase metallicity, respectively, further confirming the correlation between stellar mass and metallicity. \par
\begin{figure}
   \resizebox{\hsize}{!}
            {\includegraphics[width=20cm]{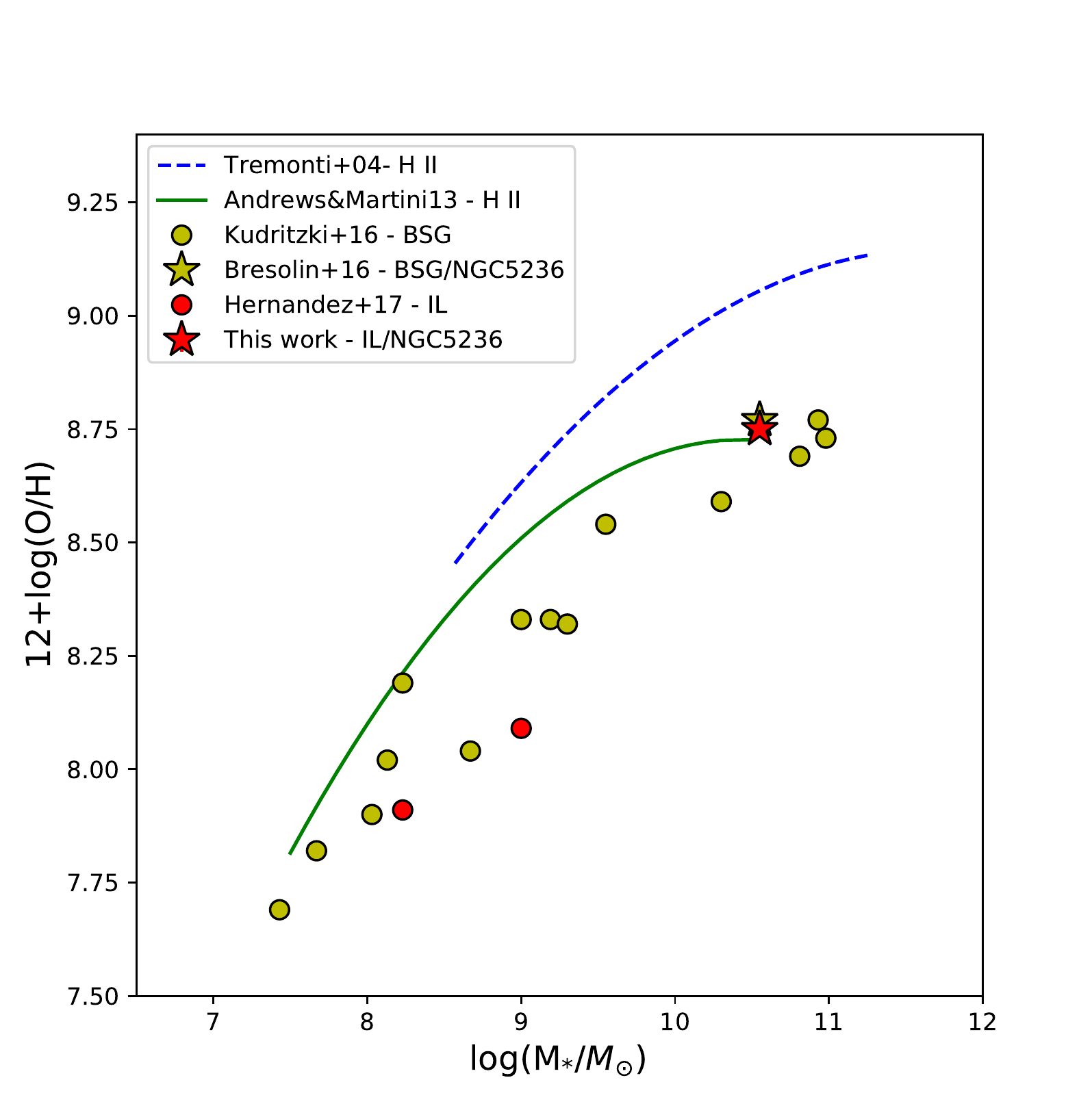}}
      \caption{Mass-metallicity relation. The dashed blue line shows the polynomial fit determined by ~\citet{tre04}. The solid green line displays the relation defined by ~\citet{and13}. The red star corresponds to the integrated metallicity for NGC 5236 obtained as part of this work. The yellow star shows the abundance estimated by \citet{bre16} for NGC 5236 using the BSG method.Yellow circles represent the stellar metallicities inferred by the BSG method, compiled by \citet{kud16}. Red circles represent the metallicities for NGC1313 and NGC1705 inferred by \citet{her17}. }
         \label{fig:mzr}
   \end{figure}

 
The MZR of star-forming galaxies has been studied exclusively through the analysis of nebular spectra. To compare the stellar and gaseous metallicity measurements, we plot our results in the mass-metallicity plane in Figure ~\ref{fig:mzr}. In this Figure we include the MZR inferred by \citet{tre04} and \citet{and13} using the Sloan Digital Sky Survey (SDSS), with dashed blue and solid green lines, respectively. Additionally, we include \citet{kud16} compilation of metallicity measurements obtained through the blue supergiant method as yellow circles and through the integrated-light method (IL) from \citet{her17} as red circles.\par

We remark that our work measures the overall metallicity of the individual clusters, [Z]. In general for spiral galaxies with metallicity gradients one adopts a characteristic metallicity measured at 0.4 R$_{25}$. According to \citet{zar94} and \citet{mou06}, metallicities of spiral galaxies at this radial distance from the center coincides with the integrated metallicity of the whole galaxy. We note that while our spectral fit analysis uses \citet{gre98}, in the following exercise we adopt the solar oxygen abundance of \citet{asp09}, 12$+$$\log$(O/H) = $8.69$. We average the metallicities measured for NGC 5236-1389, NGC 5236-1471 and NGC 5236-245 (all three YMCs located at R $\sim$ 0.4 R$_{25}$) and infer an average oxygen abundance of 12+log(O/H) = 8.75 $\pm$ 0.08 dex. \par

Using the recent stellar mass estimates of log($M_{*}/M_{\odot}$) = 10.55 by ~\citet{bre16}, our integrated metallicity for NGC 5236 is displayed as a red star, which can be compared to the metallicity for this same galaxy inferred by \citet{bre16} shown with a yellow star in Figure ~\ref{fig:mzr}. The agreement between these two measurements obtained with independent methods shows the consistency of stellar studies. \par
A compilation of stellar metallicities obtained using the BSG method  along with those using the IL method  shows that this ``stellar" MZR is rather similar to the nebular MZR inferred by ~\citet{and13} with an additional scatter and offset towards lower values. We note that the sample size of the stellar metallicity is considerably smaller than the nebular sample. Figure ~\ref{fig:mzr} supports the idea that the correlation between mass and metallicity can in principle be studied through the galactic stellar component. However, we point out that a larger measurement sample is needed to draw firmer conclusions. \par
One possible advantage of the stellar over the nebular MZR is the fact that analysis on stellar spectroscopy is more feasible on higher metallicity environments, a regime where measurements become more challenging for H II regions, especially using the direct method ~\citep{sta05,bre05,gaz15}.

\subsection{Comparison to other stellar abundances in NGC 5236} \label{comp}
\begin{figure}
   \resizebox{\hsize}{!}
            {\includegraphics[width=11.2cm]{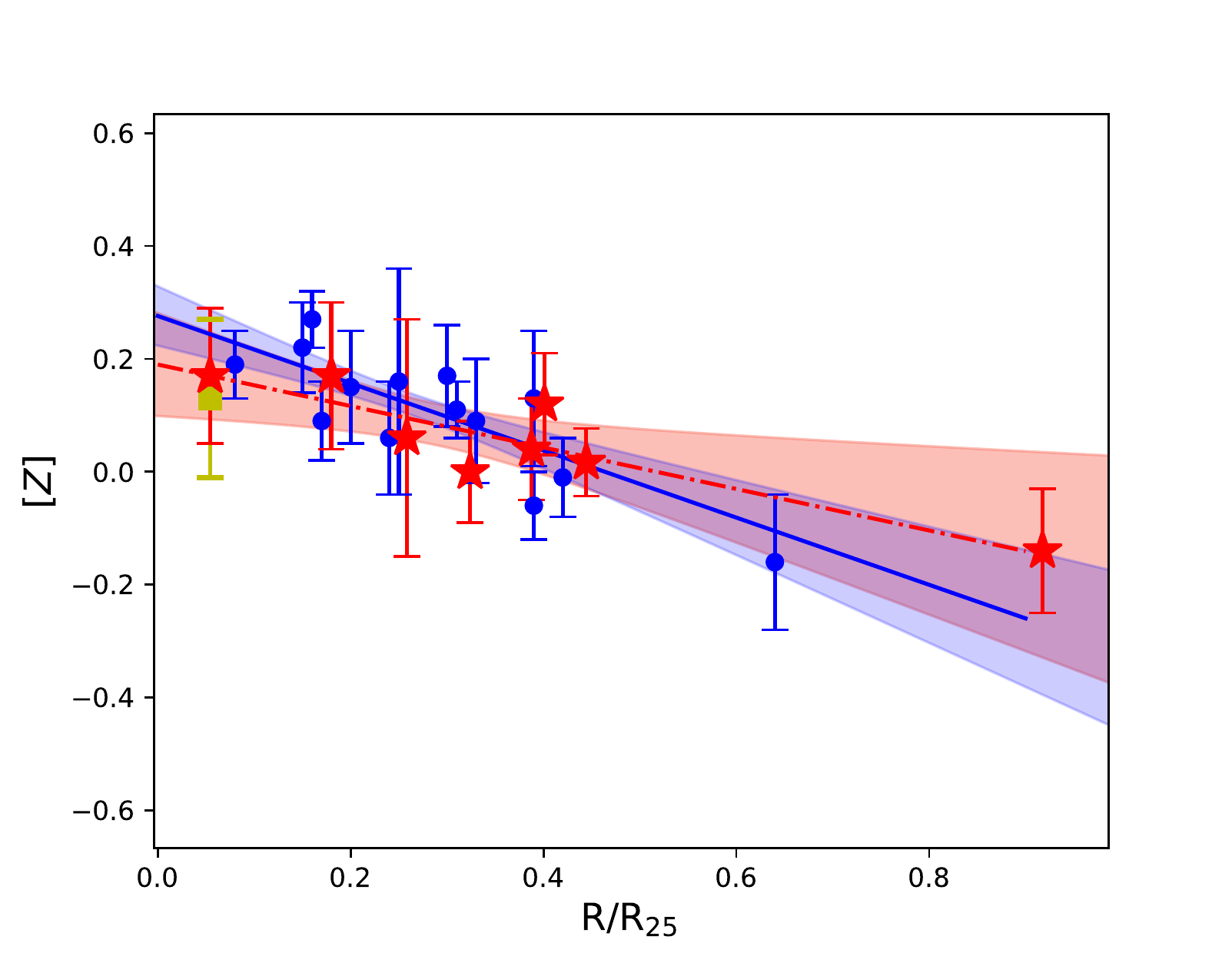}}
      \caption{Metallicities as a function of galactocentric distance normalised to isophotal radius. In red stars we display the YMC metallicity measurements obtained as part of this work. We show the metallicity measurement of ~\citet{gaz14} for NGC 5236-805 as a yellow square and in blue circles we include ~\citet{bre16} BSG metallicities. Red dashed and blue solid lines display a first-order polynomial fit for YMCs and BSGs, respectively. The salmon (YMCs) and blue (BSGs) shaded regions illustrate the 1-$\sigma$ uncertainties of the linear regressions.}
         \label{fig:z}
   \end{figure}

In Figure ~\ref{fig:z} we show the metallicities obtained as part of this work (in red stars) as a function of galactocentric distance. The galactocentric distance is normalised to the isophotal radius. This distance, R/R$_{25}$, is calculated adopting the following parameters: R$_{25}$ = 6.44 arcmin ~\citep{dev91}, $i$ = 24 deg, and PA = 45 deg ~\citep{com81}. We also include the blue supergiant metallicity measurements of ~\citet{bre16} for comparison (in blue circles), along with the YMC metallicity from ~\citet{gaz14} shown as a yellow square. \par
Before comparing the different metallicity measurements we homogenize the different sets to a single abundance scale. In this YMC work we use \citet{gre98} solar composition with a metallicity mass fraction of $Z_{\rm YMC} = 0.0169$. In contrast, \citeauthor{bre16} use \citet{asp09} solar oxygen abundance and the solar composition of \citet{gre98} for the rest of the elements with a total metallicity mass fraction $Z_{\rm BSG} = 0.0149$. We scale the BSG metallicities using the following relation
\begin{equation}
[Z]_{\rm YMC} = [Z]_{\rm BSG} - \log\bigg(\frac{Z_{\rm YMC}}{Z_{\rm BSG}}\bigg) = [Z]_{\rm BSG} - 0.06
\end{equation}
  
Using the J-band spectral analysis method ~\citet{gaz14} determined the metallicity of YMC NGC 5236-805, also included in our sample. The authors inferred a metallicity [Z] = $+$0.28 $\pm$ 0.14 dex. We point out that the study of \citet{gaz14} applies a spectral synthesis analysis based on MARCS models, which adopt solar abundances from \citet{gre07} and determine the metallicities using several elements such as Fe, Ti, Si and Mg. To account for the difference in solar abundance used in the work of ~\citeauthor{gaz14}, we revise this value considering the metallicity mass fraction of \citet{gre07}, $Z_{\rm Jband} = 0.012$, and following the relation below 
\begin{equation}
[Z]_{\rm YMC} = [Z]_{\rm Jband} - \log\bigg(\frac{Z_{\rm YMC}}{Z_{\rm Jband}}\bigg) = [Z]_{\rm Jband} - 0.15
\end{equation}
We revise the metallicity measurement by \citeauthor{gaz14} to  [Z] = $+$0.13 $\pm$ 0.14 dex. With a galactocentric distance of R/R$_{25}\sim$ 0.05, NGC 5236-805 is the inner most YMC in our work. We measure an overall metallicity of [Z] = $+$0.17 $\pm$ 0.12 for the same YMC.  This value is consistent within the errors with the J-band measurement by ~\citet{gaz14}. \par

To further explore the central stellar metallicity in NGC 5236, we compare our NGC 5236-805 metallicity with that derived by ~\citet{bre16} for a blue supergiant with a galactocentric distance of R/R$_{25}\sim$ 0.08, relatively close to our central YMC. ~\citeauthor{bre16} measure a metallicity of [Z] = $+$0.25 $\pm$ 0.06 dex, well within the errors of our inferred value. These three independent measurements, using distinct methods, show excellent agreement, confirming the above-solar metallicity environment in the central regions and inner disk of NGC 5236 and the consistency of stellar metallicities. \par
From Figure ~\ref{fig:z} we can find a strong agreement between the metallicities of blue supergiants and those of the YMCs, especially at R/R$_{25}$<0.5. In this same figure we show linear regressions to the YMC metallicities (in red dashed line) and to the blue supergiant metallicities (in blue line). We apply a linear regression only to metallicities with galactocentric distances of R/R$_{25}$<0.5, obtaining

\begin{equation}\label{eq:gradientymc}
[Z]_\mathrm{YMC} = -0.37\: (\pm 0.29) \:R/R_{25} + 0.19\: (\pm 0.09)
\end{equation}

\noindent and

\begin{equation}\label{eq:gradientbsg}
[Z]_\mathrm{BSG} = -0.60\: (\pm 0.19) \: R/R_{25} + 0.20\: (\pm 0.05)
\end{equation}

\noindent where [Z]$_\mathrm{YMC}$ applies to the YMC observations and [Z]$_\mathrm{BSG}$ to the blue supergiants. The different slopes inferred through the two methods agree within the errors of each other, with the YMC measurements having a slightly shallower gradient. We point out that gradients of $\sim-$0.4 dex R$_{25}^{-1}$ are typical for spiral galaxies \citep{ho15}. However, the gradient value inferred from the YMCs comes with large uncertainties and a flat distribution with zero gradient is well within 2$\sigma$. \par

Beyond R/R$_{25} \sim$0.5, both studies have a single metallicity measurement at different radii. In our work the YMC is at a larger galactocentric distance than the one from ~\citet{bre16}. Due to the extremely limited number of measurements beyond R/R$_{25} \sim$ 0.5, it becomes especially challenging to draw firmer conclusions regarding the spatial distribution of metallicity at larger distances from the center. Additional metallicity measurements of targets at R/R$_{25}$ > 0.5 will help discriminating between an optimal linear fit of a single or multiple gradients. \par
%Testing the full galactocentric coverage, if we instead include the full R/R$_{25}$ range in the linear regression, the slope from the YMC observations does not change significantly, going from  $-$0.37 $\pm$ 0.41  to $-$0.36 $\pm$ 0.37. For the blue supergiant slope we also see a slight change, however in the other direction, going from $-$0.60 $\pm$ 0.20 to a slightly steeper slope of $-$0.67 $\pm$ 0.18. \par

\subsection{Stellar vs. Gas Abundance}\label{stelvsgas}
The systematic offsets in the inferred metallicities using different nebular diagnostics has been discussed and studied extensively \citep[e.g.][]{ken03,mou10, lop12}. In this section we compare our stellar metallicities to those obtained through the analysis of nebular regions. We take the emission fluxes published by \citet{bre05} and estimate strong-line abundances applying the O2N2= [N II] $\lambda$6584/[O II] $\lambda$3727 method and adopting two different calibrations based on theoretical models and empirical data. We mainly focus on the metallicity characterisation of the inner disc (R/R$_{25}$ < 0.6) of NGC 5236. We note that there are several other strong-line diagnostics which we have not included here.  \citet{bre16} provide a detailed discussion on strong-line diagnostics along with an extensive comparison to their predicted chemical abundances. The aim of this section is to understand how the stellar metallicities obtained in this work compare to the general trends and values of nebular studies in a general sense and how much our results resemble or differ from those obtained by \citet{bre16}. \par

\textbf{O2N2 - Theoretical:} We apply the strong-line calibration for O2N2 by \citet{kew02}. We refer to this calibration as K02. The method is calibrated using theoretical photoionisation models. 

\textbf{O2N2 - Empirical:} The O2N2 method by \citet{bre07-2} is based on a sample of 140 direct abundance measurements from extragalactic H II regions. We refer to this calibration as B07

\begin{figure}
   \resizebox{\hsize}{!}
            {\includegraphics[width=11.2cm]{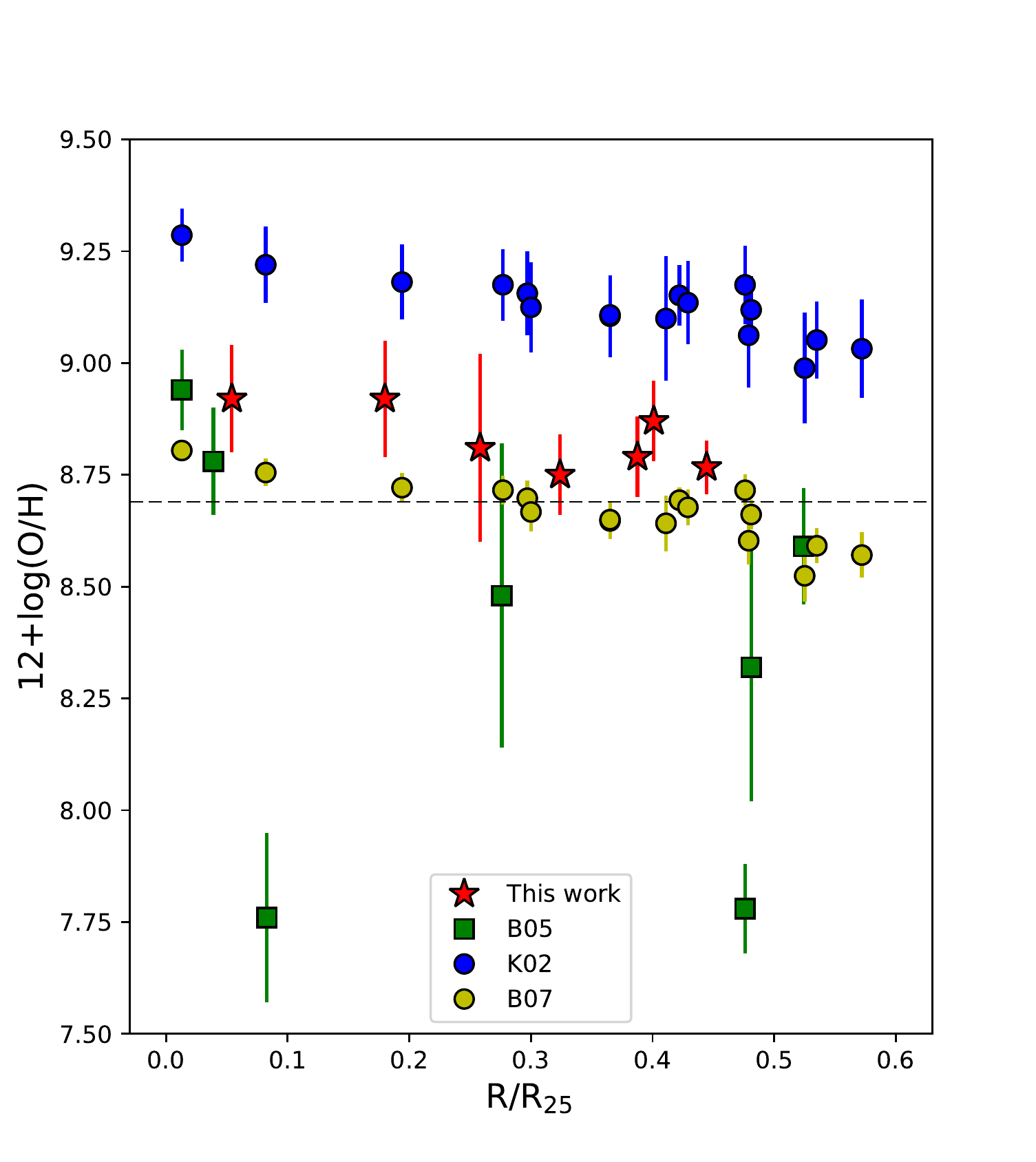}}
      \caption{Oxygen abundance as a function of galactocentric distance normalized to the isophotal radius. In red stars we show the metallicity measurements converted to oxygen abundance inferred in this work. In blue circles we show the oxygen measurements from the O2N2 calibration by K02. In yellow circles we show the abundances from B07. In green squares we show the oxygen abundances inferred from the direct method by B05.}
         \label{fig:stellar_vs_gas}
   \end{figure}

In addition to comparing the metallicities presented in this work to the nebular calibrations above, we also include the abundances obtained by \citet{bre05} using the direct method. We refer to these measurements as B05. In Figure ~\ref{fig:stellar_vs_gas} we show the oxygen abundances using these four different methods: O2N2/Theoretical (K02), O2N2/Empirical (B07), direct method (B05) and IL (this work).  A visual inspection of this Figure shows rather similar slopes for the stellar (in red stars), K02 (in blue circles) and B07 (in yellow circles). On the other hand \citet{bre16} find that all the strong-line indicators they investigate, including K02 and B07, have shallower slopes than those measured from the BSG abundances. Considering we find similarities between our slopes and those from K02 and B07, this difference between the gradient by \citet{bre16} and those from strong-line indicators (K02 and B07) is expected from the inferred gradients for YMCs and BSG shown in Eq. ~\ref{eq:gradientymc} and ~\ref{eq:gradientbsg}, where we see that YMCs point at a shallower slope. Furthermore, a clear offset is present where the oxygen abundances from K02 are higher than our YMC work, by $\sim$0.3-0.4 dex, and the B07 abundances, by $\sim$0.5 dex. In this context, these results are similar to those observed in \citet{bre16} with the BSG abundances lying $\sim$ 0.2-0.3 dex lower than those calibrated with the K02 method. \par
The abundances from the direct method, B05, exhibit a rather strong scatter, however, the inner most measurements agree well with our stellar metallicities. We point out that the abundance from B05 of 12+log(O/H)=7.75 at R/R$_{25}$$\sim$0.08 is merely a lower limit. At R/R$_{25}$ > 0.2 the B05 abundances deviate from ours to lower values.\par
In this comparison, the best agreement between the nebular and stellar abundances is obtained from the empirical B07 calibration, although our measurements are consistently higher than those from the B07 diagnostic. Linear regressions for the B07 data and our measurements show consistent slopes, with our metallicities being offset to higher metallicities by $\sim$0.1 dex. As pointed out in \citet{bre16}, in these comparisons we do not account for the effect of oxygen depletion (e.g. onto interstellar dust grains). This effect is especially important for oxygen abundances that have been derived from empirical calibrations, such as B07. \citet{mes09} and \citet{pei10} have empirically determined depletion factors ranging between $-$0.08 and $-$0.12 dex. Applying an average correction for $-$0.1 dex of depletion to the B07 nebular abundances brings the measurements to better agreement with our YMC stellar abundances.\par

\subsection{Comparison to Chemical Evolution Models} \label{comp_mod}
We now compare our direct metallicity measurements with chemical evolution models produced specifically for NGC 5236. \par
\citet{bre16} introduced two chemical evolution models for their observed present-day metallicity distribution over the entire NGC 5236 disc. For details on the construction of the different models we refer the reader to \citet{bre16}. Briefly, their individual models were generated using the analytical chemical evolution model of \citet{kud15}. This analytical model improves over the closed-box scenario \citep{pag75} by accounting for the influence of gas flows (in and out) to regulate the spatial distribution of abundances. The model of \citeauthor{kud15} provides theoretical radial metallicity distributions based on specified stellar and gas radial mass profiles with two additional free parameters, infall and outflow. To generate a closed-box model, these two free parameters are set to 0.\par
\begin{figure}
   \resizebox{\hsize}{!}
            {\includegraphics[width=11.2cm]{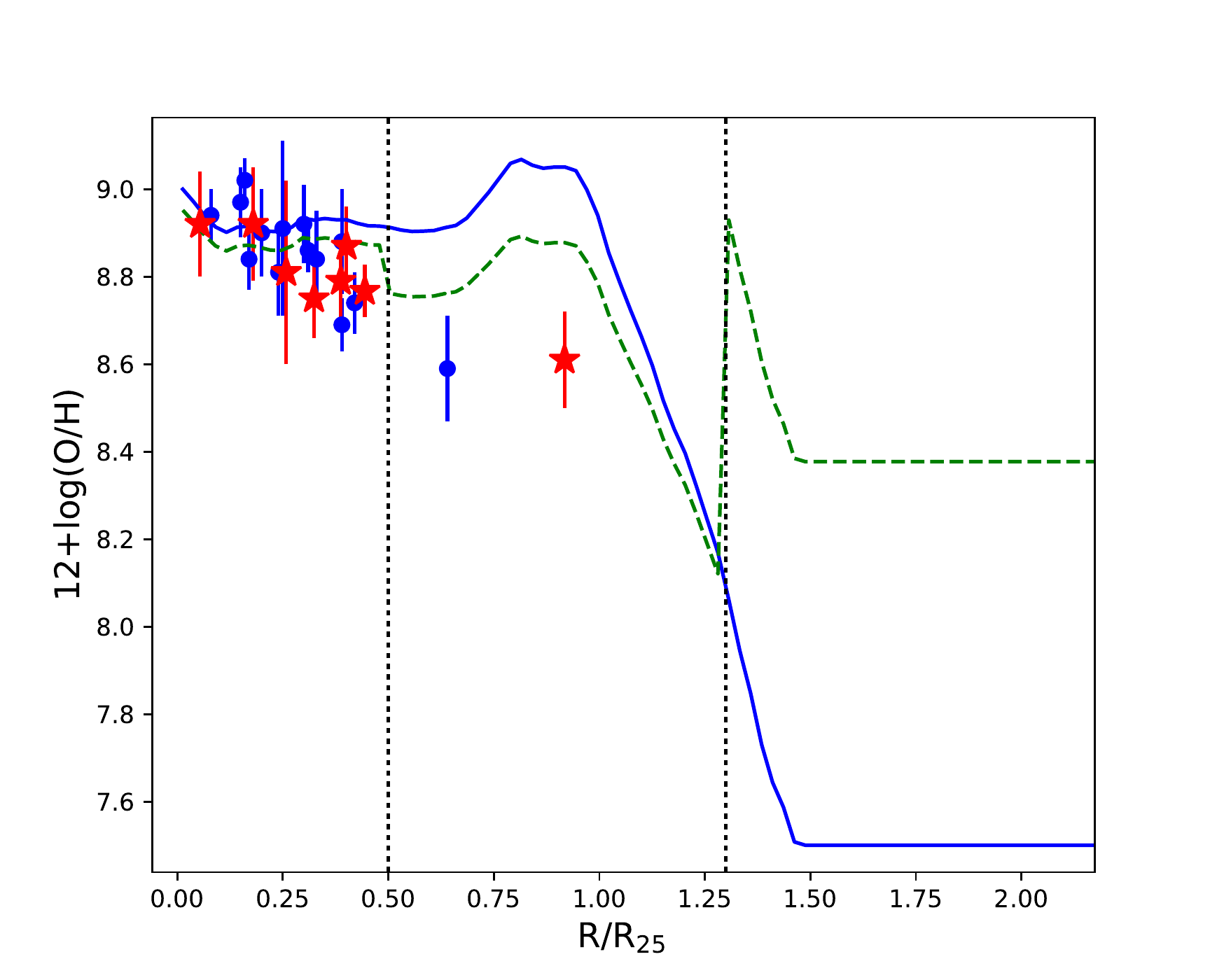}}
      \caption{Oxygen abundance as a function of galactocentric distance normalised to isophotal radius. In red stars we show the YMC measurements inferred in our analysis. In blue circles we include BSG oxygen abundances from \citet{bre16}. The blue and green dashed lines represent two chemical evolution models for NGC 5236 \citep{bre16}, accounting for galactic winds (and infall) and closed-box, respectively. The dashed vertical lines show the radial divisions used in generating the detailed model. }
         \label{fig:chemod}
   \end{figure}

For the case of the detailed model involving galactic winds, and in/outfalls, the radial range (R/R$_{25}$) was divided in three different sections (0.0-0.5, 0.5-1.3, and 1.3-1.5) where the authors vary the mass flow rates and the infalling gas. The infall and outfall parameters are defined as the ratio of mass infall/outfall rate by the star formation rate, $\dot{M}_{acct}$/$\psi$ and $\dot{M}_{loss}$/$\psi$. The best model fit (shown in Figure ~\ref{fig:chemod}) required $\dot{M}_{acct}$/$\psi$ = 0.0 and $\dot{M}_{loss}$/$\psi$ = 0.12 for the first section,  $\dot{M}_{acct}$/$\psi$ = 0.0 and $\dot{M}_{loss}$/$\psi$ = 0.50 for the second region, and $\dot{M}_{acct}$/$\psi$ = 1.0 and $\dot{M}_{loss}$/$\psi$ = 0.0 for the outer disk. \par

In Figure~\ref{fig:chemod} we show the detailed (infall$+$galactic winds) and closed-box chemical models by ~\citet{bre16}, along with our abundance measurements and those from BSGs. We convert our overall metallicities measured in NGC 5236 adopting a solar oxygen abundance of 12+log(O/H)=9.69 from \citet{asp09}. We note that the feature in Figure ~\ref{fig:chemod} in the detailed model (green dashed line) around R/R$_{25} \sim$ 1.3 is an artificial spike originating from the two connecting radial sections described above. For distances R/R$_{25}$<0.5 both models predict relatively similar abundance gradients, although the closed-box model gives slightly higher values. \par
In general our observed abundances agree slightly better with the lower oxygen abundances predicted by the detailed model (green dashed line), mainly found in the first radial region where no gas infall is required, and only a small fraction of the material is expelled due to galactic winds. Furthermore, it is clear that for our last abundance measurement at R/R$_{25}$ = 0.91, the detailed model predicts a value closer to our oxygen abundance than the closed-box model. Based on this detailed model and our abundance measurement in this second radial region it appears reasonable to assume a different gradient to describe the metallicity distribution in this region of the disk. However, more stellar metallicity measurements are needed to verify this statement.\par

\subsection{Metallicity-Age Relation} \label{mar}
\begin{figure}
   \resizebox{\hsize}{!}
            {\includegraphics[width=11.2cm]{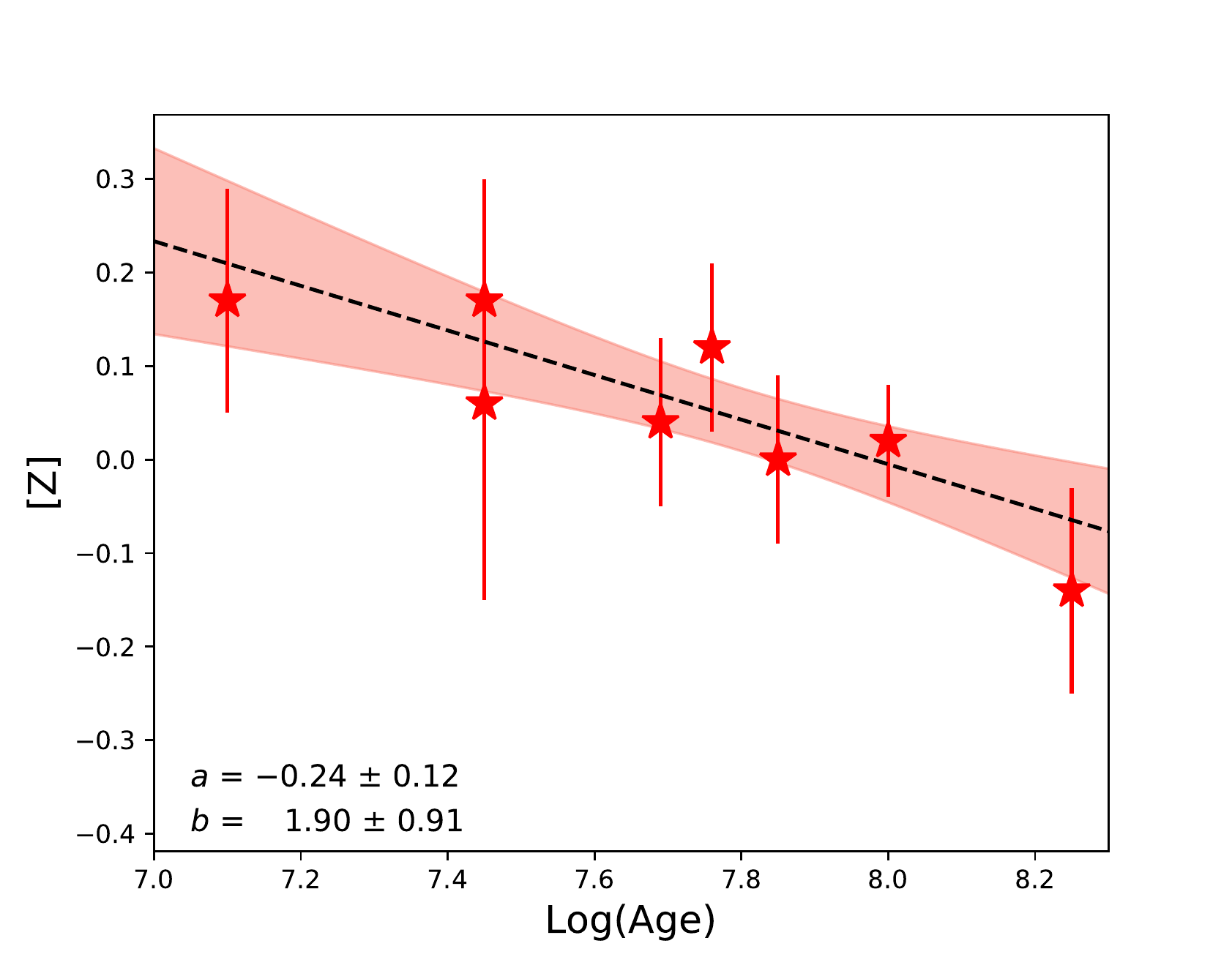}}
      \caption{Metallicity as a function of Log(age). In red stars we show the metallicity measurements obtained as part of this work. We show a first order polynomial fit as a black dashed line. In the shaded salmon region we show 1$\sigma$ confidence intervals. We include the slope ($a$) and zero-point ($b$) of the linear regression along with their uncertainties. }
         \label{fig:mar}
   \end{figure}

We observe a clear anticorrelation between the measured metallicities and their corresponding ages. In Figure ~\ref{fig:mar} we show this relation along with a first-order polynomial fit of the form [$Z$] = $a$ Log(age) $+$ $b$, represented by a black dashed line. 
%We find a $\chi_{min}^{2}$ = 1.8 and determine that the probability of observing a value equal or larger than this $\chi_{min}^{2}$ is $P(\chi^{2} \geq \chi_{min}^{2})$ = 0.94. To assess if this anitcorrelation is significant, we estimate the $\chi^{2}$ for a zero-slope scenario ($a$ = 0), and we find $\chi_{\rm zero}^{2}$ = 6.0. The probability of observing this $\chi_{\rm zero}^{2}$ or larger is then $P(\chi^{2} \geq \chi_{\rm zero}^{2})$ = 0.43. 
We estimate a slope of $a = -0.24 \pm 0.12$, with a $2$-$\sigma$ correlation hinting at a minimum decline in metallicity of $\sim$0.1 dex in a time period of $\sim$100 Myr.
We note that the oldest YMCs in our sample, NGC5236-254, has a location R/R$_{25}$>0.5 and a metallicity lower than the rest by $\sim$0.15 dex. Similarly one of the youngest clusters, NGC5236-805, is the most centrally located and one of the most metal rich objects in our study. While these observations suggest that the anticorrelation could in principle be of a chemical evolution origin, we can not discard systematic effects in the spectral fitting as a possible cause. 

%______________________________________________________________

\section{Conclusions}\label{con}
Chemical abundances of star-forming galaxies, especially beyond the Local Group, are mainly based on the analysis of nebular emission lines. A characteristic problem of nebular studies arises when comparing the abundances obtained through the different calibrators (e.g. O2N2, O3N2, N2) where one can find systematic offsets as high as $\sim$0.7 dex \citep{bre08,kew08}. To avoid these poorly understood systematic uncertainties, in this paper we carry out a stellar metallicity analysis on a sample of eight extragalactic YMCs distributed throughout NGC 5236. This stellar abundance approach is of special relevance for environments of metallicities above solar, where certain nebular methods fail or tend to underestimate abundances \citep{sta05,zur12,sim11}. \par
We apply the abundance technique developed by L12 for integrated-light observations and show that this can be successfully used on intermediate-resolution spectroscopic data taken with the X-shooter spectrograph of objects in the high metallicity range. We derive precise metallicities and find excellent agreement with independent stellar metallicity studies in NGC 5236. We measure a super-solar metallicity of [Z] = $+$0.17 $\pm$ 0.12 dex for the most centrally located YMC NGC 5236-805. \par %Additionally, we observe a metallicity gradient in the central parts of the galactic disk (R/R$_{25}$<0.05), and speculate that this gradient could be changing at larger radii (0.05<R/R$_{25}$<1.0) but require additional metallicity measurements to draw firmer conclusions. \par
We further compare our abundance measurements to chemical evolution models by \citet{bre16}. Similar to their findings, we observe that their best model, which accounts for galactic winds and in/outflows, reproduces our observed abundances better than their simple closed-box model. Based on this comparison we conclude that the central regions of NGC 5236 are possibly experiencing no infall of material, and a small loss of material due to galactic winds. \par
We conclude that the analysis of integrated-light observations is an independent and reliable method for obtaining metallicities and studying galactic abundance gradients in star forming galaxies in high metallicity environments. Our results also prove that the X-Shooter spectrograph allows for these type of abundance studies today and expect future instrumentation and telescopes such as the Extremely Large Telescope (ELT), the Giant Magellan Telescope (GMT), and the Thirty Meter Telescope (TMT) to continue providing essential information on the chemical enrichment of other galaxies. Furthermore, the excellent agreement between two independent methods, IL and BSGs, is especially encouraging for future work with these new generation telescopes as an alternative to \ion{H}{ii}-techniques allowing us to expand our knowledge of galaxy formation and evolution. \par

\section*{Acknowledgements}
We thank A. Gonneau, Y.-P. Chen and M. Dries for their help and guidance during the X-Shooter reduction process. We are especially thankful to R.-P. Kudritzki  for his detailed review of this manuscript which improved the quality of our work. This research has made use of the NASA/IPAC Extragalactic Database (NED), which is operated by the Jet Propulsion Laboratory, California Institute of Technology, under contract with the National Aeronautics and Space Administration.

%%%%%%%%%%%%%%%%%%%%%%%%%%%%%%%%%%%%%%%%%%%%%%%%%%



%\bibliographystyle{mnras}


%%%%%%%%%%%%%%%%%%%%%%%%%%%%%%%%%%%%%%%%%%%%%%%%%%

%%%%%%%%%%%%%%%%% APPENDICES %%%%%%%%%%%%%%%%%%%%%

\appendix
\clearpage

\section{Metallicities as a function of wavelength}
We present tables displaying the individual bin measurements for each of the YMCs studied in this work. 

\begin{table}
\caption{Metallicities for NGC 5236-245}
\label{z245}
 \centering 
\begin{tabular}{ccc} 
 \hline  \hline \
$Wavelength\, [$\AA$]$& [Z] &Error\\
  \hline
4000-4200& $+$0.043& 0.265  \\
4200-4400& $+$0.236& 0.454  \\
6100-6300& $-$0.374& 0.129\\
6300-6500& $+$0.199& 0.127\\
6588-6700& $+$0.069& 0.129 \\
6700-6800& $-$0.183& 0.192 \\
7400-7550& $+$0.003& 0.093 \\
8500-8700& $+$0.034& 0.065\\
8700-8830& $+$0.129& 0.102  \\
 \hline 
 \end{tabular}
\end{table}

\begin{table}
\caption{Metallicities for NGC 5236-254}
\label{table:z254}
 \centering 
\begin{tabular}{ccc} 
 \hline  \hline \
$Wavelength\, [$\AA$]$& [Z] &Error\\
  \hline
4000-4200& $+$0.133& 0.134\\
4200-4400& $+$0.133& 0.104\\
6100-6300& $+$0.153& 0.087\\
6300-6500& $-$0.175& 0.136\\
6588-6700& $-$0.502& 0.175\\
6700-6800& $-$0.573& 0.242 \\
7400-7550& $+$0.369& 0.100\\
8500-8700& $-$0.305& 0.043\\
8700-8830& $-$0.400& 0.104\\
 \hline 
 \end{tabular}
\end{table}

\begin{table}
\caption{Metallicities for NGC 5236-367}
\label{table:z367}
 \centering 
\begin{tabular}{ccc} 
 \hline  \hline \
$Wavelength\, [$\AA$]$& [Z] &Error\\
  \hline
4000-4200& $-$0.641& 0.129 \\
4200-4400& $-$0.271& 0.132\\
6100-6300& $-$0.175& 0.071\\
6300-6500& $+$0.199& 0.074  \\
6588-6700& $-$0.008& 0.112 \\
6700-6800& $-$0.180& 0.159 \\
7400-7550& $+$0.159& 0.075\\
8500-8700& $-$0.192& 0.130\\
8700-8830& $+$0.269& 0.081\\
 \hline 
 \end{tabular}
\end{table}

\begin{table}
\caption{Metallicities for NGC 5236-805}
\label{table:z805}
 \centering 
\begin{tabular}{ccc} 
 \hline  \hline \
$Wavelength\, [$\AA$]$& [Z] &Error\\
  \hline
4000-4200& $+$0.361& 0.033 \\
4200-4400& $+$0.026& 0.047\\
6100-6300& $+$0.039& 0.027\\
6300-6500& $+$0.090& 0.033 \\
6588-6700& $+$0.155& 0.029 \\
6700-6800& $-$0.116& 0.053 \\
7400-7550& $-$0.284& 0.126\\
8500-8700& $+$0.951& 0.113\\
8700-8830& $+$0.292& 0.024\\
 \hline 
 \end{tabular}
\end{table}

\begin{table}
\caption{Metallicities for NGC 5236-1182}
\label{table:z1182}
 \centering 
\begin{tabular}{ccc} 
 \hline  \hline \
$Wavelength\, [$\AA$]$& [Z] &Error\\
  \hline
4000-4200& $+$0.302& 0.047 \\
4200-4400& $+$0.312 & 0.035 \\
6100-6300& $+$0.406& 0.043\\
6300-6500& $-$0.344& 0.065 \\
6588-6700& $-$0.446& 0.051 \\
6700-6800& $-$0.548& 0.073 \\
7400-7550& $+$0.188& 0.026\\
8500-8700& $-$0.426& 0.108\\
8700-8830& $+$0.251& 0.021 \\
 \hline 
 \end{tabular}
\end{table}

\begin{table}
\caption{Metallicities for NGC 5236-1234}
\label{table:z1234}
 \centering 
\begin{tabular}{ccc} 
 \hline  \hline \
$Wavelength\, [$\AA$]$& [Z] &Error\\
  \hline
4000-4200& $-$0.034& 0.104 \\
4200-4400& $-$0.764& 0.149 \\
6100-6300& $-$0.938& 0.171\\
6300-6500& $-$0.816& 0.109 \\
6588-6700& $+$0.863& 0.096 \\
6700-6800& $+$0.053& 1.038 \\
7400-7550& $+$0.263& 0.055 \\
8500-8700& $-$0.732& 0.132\\
8700-8830& $+$0.177& 0.053 \\
 \hline 
 \end{tabular}
\end{table}

\begin{table}
\caption{Metallicities for NGC 5236-1389}
\label{table:z1389}
 \centering 
\begin{tabular}{ccc} 
 \hline  \hline \
$Wavelength\, [$\AA$]$& [Z] &Error\\
  \hline
4000-4200& $-$0.197& 0.083 \\
4200-4400& $-$0.506& 0.073 \\
6100-6300& $-$0.342& 0.165\\
6300-6500& $-$0.003& 0.066 \\
6588-6700& $+$0.008& 0.140 \\
6700-6800& $+$0.064& 0.103 \\
7400-7550& $+$0.310& 0.052\\
8500-8700& $-$0.400& 0.125\\
8700-8830& $+$0.228& 0.048 \\
 \hline 
 \end{tabular}
\end{table}

\begin{table}
\caption{Metallicities for NGC 5236-1471}
\label{table:z1471}
 \centering 
\begin{tabular}{ccc} 
 \hline  \hline \
$Wavelength\, [$\AA$]$& [Z] &Error\\
  \hline
4000-4200& $-$0.331& 0.124 \\
4200-4400& $+$0.342& 0.091 \\
6100-6300& $+$0.016& 0.108\\
6300-6500& $+$0.246& 0.100 \\
6588-6700& $+$0.326& 0.204 \\
6700-6800& $+$0.008& 0.142 \\
7400-7550& $+$0.256& 0.061\\
8500-8700& $-$0.082& 0.032\\
8700-8830& $+$0.568& 0.050\\
 \hline 
 \end{tabular}
\end{table}

%%%%%%%%%%%%%%%%%%%%%%%%%%%%%%%%%%%%%%%%%%%%%%%%%%

% Don't change these lines
\bsp	% typesetting comment
\label{lastpage}
\end{document}